\documentclass{ECOC-author}

\usepackage{lipsum}
\usepackage{subcaption}
\newcommand{\vect}[1]{\ensuremath{\boldsymbol{#1}}}
\newcommand{\mat}[1]{\ensuremath{\mathbf{#1}}}
\newcommand{\imag}{\jmath}

\begin{document}

%\title{\textsc{Revisiting Multi-Step Nonlinearity Compensation with
%Insights from Machine Learning}}

%\title{\textsc{Revisiting Multi-Step with Machine Learning}}

%\title{\textsc{Revisiting Multi-Step Nonlinearity Compensation with Machine Learning}}

\title{{REVISITING MULTI-STEP NONLINEARITY COMPENSATION WITH MACHINE LEARNING}}

%\title{PAPER TITLE}
%\title{\textsc{Paper Title}}

%\title{REVISITING MULTI-STEP NONLINEARITY COMPENSATION WITH
%MACHINE LEARNING}

\author{Christian H\"ager\ad{1,2}\corr, Henry D~Pfister\ad{2}, Rick
M~B\"utler\ad{3}, Gabriele Liga\ad{3}, Alex Alvarado\ad{3}}

% (use style: unabbreviated first name, middle name initials, last name e.g. John A K Smith\ad{1}, Edward Jones\ad{2})

\address{\add{1}{Department of Electrical Engineering, Chalmers
University, Gothenburg, Sweden}
\add{2}{Department of Electrical and Computer Engineering, Duke
University, Durham, USA}
\add{3}{Department of Electrical Engineering, Eindhoven University of
Technology, Eindhoven, The Netherlands}
\email{christian.haeger@chalmers.se}}

%Enter a maximum of {[three ECOC 2019 sub-categories /or/ five key
%words selected from
%https://journals.ieeeauthorcenter.ieee.org/create-your-ieee-article/create-the-text-of-your-article/ieee-standardized-keywords/]
%here.}

%\vspace*{-0.2cm}

\keywords{MACHINE LEARNING BASED DSP, DEEP LEARNING,
LOW-COMPLEXITY DIGITAL BACKPROPAGATION, SUBBAND PROCESSING,
POLARIZATION MODE DISPERSION}

\begin{abstract}
	For the efficient compensation of fiber nonlinearity, one of the
	guiding principles appears to be: fewer steps are better and more
	efficient. We challenge this assumption and show that carefully
	designed multi-step approaches can lead to better
	performance--complexity trade-offs than their few-step
	counterparts. 
\end{abstract}

%The abstract should be suitable for direct inclusion in abstracting
%services as a self-contained article. The length of the abstract
%should not exceed 45 words. Do not include figure numbers, table
%numbers, references or displayed mathematical expressions

\maketitle

\section{Introduction}

%, Fougstedt2017b

Mitigating fiber nonlinearity is a significant challenge in high-speed
fiber-optic communication systems. In principle, digital
backpropagation (DBP) based on the split-step Fourier method (SSFM)
offers ideal compensation of deterministic propagation impairments
including nonlinear effects. On the other hand, several authors have
highlighted the large computational burden associated with a real-time
digital signal processing (DSP) implementation and proposed various
techniques to reduce the complexity \cite{Ip2008, Du2010,
Rafique2011a, Napoli2014, Jarajreh2015, Giacoumidis2015,
Secondini2016, Fougstedt2017, Haeger2018ofc}. In many of these works,
the number of steps (or compensation stages) is used not only to
quantify complexity but also as a general measure of the quality for
the proposed complexity-reduction method. In a nutshell, one gets the
impression that fewer steps are better and more efficient. 

While previous work has indeed demonstrated that complexity savings
are possible by reducing steps, the main purpose of this paper is to
highlight the fact that fewer steps are not more efficient \emph{per
se}. In fact, recent progress in machine learning suggests that deep
computation graphs with many steps (or layers) tend to be more
parameter efficient than shallow ones that use fewer steps. In this
paper, we illustrate through various examples how this insight can be
applied in the context of fiber-nonlinearity compensation in order to
achieve low-complexity and hardware-efficient DBP. 

\section{Supervised Machine Learning}

We start by reviewing the standard supervised learning setting for
feed-forward neural networks (NNs). A feed-forward NN with $M$ layers
defines a mapping $\vect{y} = f_\theta(\vect{x})$ where the input
vector $\vect{x} \in \mathcal{X}$ is mapped to the output vector
$\vect{y} \in \mathcal{Y}$ by alternating between affine
transformations $\vect{z}^{(i)} = \vect{W}^{(i)} \vect{x}^{(i-1)} +
\vect{b}^{(i)}$ and pointwise nonlinearities $\vect{x}^{(i)} =
\phi(\vect{z}^{(i)})$ with $\vect{x}^{(0)} = \vect{x}$ and
$\vect{x}^{(M)} = \vect{y}$. The parameter vector $\theta$ comprises
all elements of the weight matrices $\vect{W}^{(1)}, \dots,
\vect{W}^{(M)}$ and vectors $\vect{b}^{(1)},\dots,\vect{b}^{(M)}$.
Given a training set $S\subset \mathcal{X} \times \mathcal{Y}$ that
contains a list of desired input--output pairs, training proceeds by
minimizing the empirical loss $\mathcal{L}_S (\theta) \triangleq
\frac{1}{|S|} \sum_{(\vect{x},\vect{y})\in S} \ell \big(
f_\theta(\vect{x}),\vect{y})$, where $\ell(\hat{\vect{y}},\vect{y})$
is the per-sample loss associated with returning the output
$\hat{\vect{y}}$ when $\vect{y}$ is correct. When the training set is
large, one typically optimizes $\theta$ using a variant of stochastic
gradient descent (SGD). In particular, mini-batch SGD uses the
parameter update $\theta_{t+1} = \theta_t - \alpha \nabla_\theta
\mathcal{L}_{B_t} (\theta_t)$, where $\alpha$ is the step size and
$B_t \subseteq S$ is the mini-batch used in the $t$-th step. 

Supervised machine learning is not restricted to NNs and learning
algorithms such as SGD can be applied to other function classes as
well. In this paper, we do not further consider NNs, but instead focus
on approaches where the function $f_\theta$ results from
parameterizing model-based algorithms. In fact, prior to the current
revolution in machine learning, communication engineers were quite
aware that system parameters (such as filter coefficients) could be
learned using SGD. It was not at all clear, however, that more
complicated parts of the system architecture could be learned as well.
For example, in the linear operating regime, polarization mode
dispersion (PMD) can be compensated by choosing the function
$f_\theta$ as the convolution of the received signal with the impulse
response of a linear multiple-input multiple-output (MIMO) filter,
where $\theta$ corresponds to the filter coefficients. For a suitable
choice of the loss function $\ell$, applying SGD then maps into the
well-known constant modulus algorithm \cite{Savory2008}. More details
and extensions of this approach are discussed in Sec.~\ref{sec:pmd}. 

%
%
%\emph{Example:} 
%
%\demo

\section{Multi-Step Digital Backpropagation}

%This approach is referred to as digital backpropagation (DBP) and was
%inspired by a similar idea where optical components were used for the
%processing \cite{Pare1996}. DBP was first studied as a transmitter
%pre-distortion technique \cite{Essiambre2005, Roberts2006}.

Signal propagation in an optical fiber is described by the nonlinear
Schr\"odinger equation (NLSE). For efficient DBP, the task is to
approximate the solution of the NLSE using as few computational
resources as possible. The SSFM computes a numerical solution by
alternating between linear filtering steps accounting for chromatic
dispersion (CD) and nonlinear phase rotation steps accounting for the
optical Kerr effect \cite[p.~40]{Agrawal2006}.  This is quite similar
to the functional form of a deep NN \cite{Haeger2018ofc}. 

\subsection{Short Chromatic-Dispersion Filters}
\label{sec:dbp}

Real-time DBP based on the SSFM is widely considered to be impractical
due to the complexity of the fast Fourier transforms (FFTs) commonly
used to implement frequency-domain (FD) CD filtering. To address this
issue, time-domain (TD) filtering with finite impulse response (FIR)
filters has been suggested in, e.g., \cite{Ip2008, Zhu2009,
Goldfarb2009, Fougstedt2017}. In these works, either a single filter
or filter pair is designed and then used repeatedly in each step.
However, using the same filter multiple times is suboptimal in general
and the filter coefficients in the entire DBP algorithm should be
optimized jointly. To that end, one can use supervised learning based
on SGD by letting the function $f_\theta$ be the SSFM, where the
linear steps are implemented with FIR filters and $\theta$ corresponds
to the filter coefficients used in \emph{all} steps
\cite{Haeger2018ofc, Haeger2018isit}. 

As an example, consider single-channel DBP of a $10$-Gbaud signal over
$25 \times 80$ km of standard single-mode fiber using the SSFM with
one step per span. For this scenario, Ip and Kahn have shown that
$70$-tap filters are required to obtain acceptable accuracy
\cite{Ip2008}. This assumes that the filters are designed using FD
sampling and that the same filter is used in each step. The resulting
hardware complexity was estimated to be over $100$ times larger than
for linear equalization. On the other hand, with jointly optimized
filters, it is possible to achieve similar accuracy by alternating
between filters that are as short as $5$ and $3$ taps
\cite{Haeger2018isit}. This reduces the complexity by almost two
orders of magnitude, making it comparable to linear equalization.
Fig.~\ref{fig:filter} shows the DSP architecture assuming a folded
implementation that takes advantage of filter symmetry. 

\setcounter{figure}{1}
\begin{figure*}[t]
	\centering
	\includegraphics{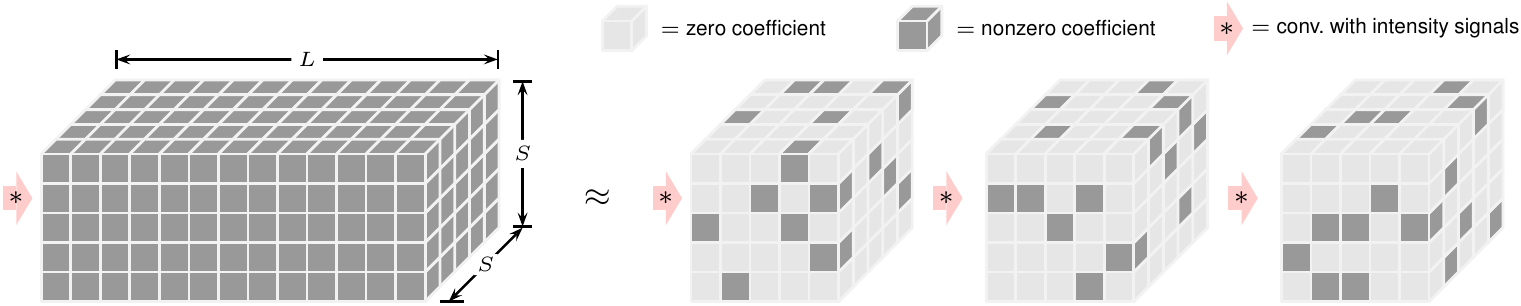}

	\caption{Tensor representation of an $L$-tap $S \times S$
	MIMO filter for DBP based on subband processing, where $S$ is the
	number of subbands (left); learned multi-step decomposition with
	sparse subfilters (right). }
	%$\mat{G}(z) = \mat{G}^{(1)}(z)\mat{G}^{(2)}(z)\mat{G}^{(3)}$(z). }
	\label{fig:subbands}
\end{figure*}

At first glance, it may not be clear why multi-step DBP can benefit
from a joint parameter optimization. After all, the standard SSFM
applies \emph{the same} CD filter many times in succession, without
the need for any elaborate optimization. The explanation is that in
the presence of practical imperfections such as finite-length filter
truncation, applying the same \emph{imperfect} filter multiple times
can be detrimental because it magnifies any weakness.  To achieve a
good combined response of neighboring filters and a good overall
response, the truncation of each filter needs to be delicately
balanced. 

%A theoretical justification for the approach can be found in
%\cite{Haeger2018isit} where the optimization problem is modeled using
%a multi-objective cost function. 

%Indeed, after examining the optimized individual (per-step)
%filter responses, we found that they are generally worse
%approximations to the ideal CD frequency response compared to, e.g.,
%least-squares optimal filters. However, the combined response of
%neighboring filters and also the overall response of all filters is
%much better compared the conventional strategy of using the same
%filter in each step. 

%In fact, using the same imperfect filter multiple times magnifies any
%weakness.  To avoid this problem and achieve a good overall response,
%different CD filters should be used at each step. 

%The same reasoning should also apply to other operators, e.g.,
%Volterra filters. 

\setcounter{figure}{0}
\begin{figure}[t]
	\centering
		\includegraphics{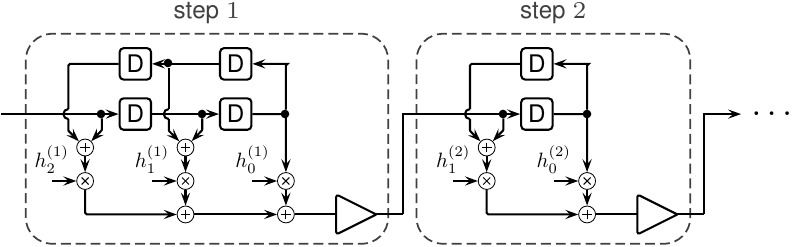}
	\caption{Folded CD filters for efficient multi-step DBP. 
	Triangles represent the (memoryless) nonlinear phase rotations.}
	\label{fig:filter}
\end{figure}

\subsection{Sparse MIMO Filters for Subband Processing}

%This can allow for significant complexity savings. 
%The idea is to split the received signal into $N$ parallel signals
%using a filter bank. Assuming a bandwidth reduction by $N$, the delay
%spread per subband signal is reduced by $N^2$. 

The complexity of DBP with TD filtering is largely dominated by the
total number of required CD filter taps in all steps and this
increases quadratically with bandwidth, see, e.g., \cite{Taylor2008,
Ho2009}. Thus, efficient TD-DBP of wideband signals is challenging.
One possible solution is to employ subband processing and split the
received signal into $S$ parallel signals using a filter bank
\cite{Taylor2008, Ho2009, Slim2013, Nazarathy2014, Mateo2010, Ip2011,
Oyama2015, Haeger2018ecoc}. A theoretical foundation for DBP based on
subband processing is obtained by inserting the split-signal
assumption $u = \sum_{i=1}^S u_i$ into the NLSE. This leads to a set
of coupled equations which can then be solved numerically. We focus on
the modified SSFM proposed in \cite{Leibrich2003} which is essentially
equivalent to the standard SSFM for each subband, except that all
sampled intensity waveforms $|u_1|^2, \ldots, |u_S|^2$ are jointly
processed with a MIMO filter prior to each nonlinear phase rotation
step. This accounts for cross-phase modulation between subbands but
not four-wave mixing because no phase information is exchanged. 

MIMO filters can be relatively demanding in terms of hardware
complexity. As an example, consider the scenario in
\cite{Haeger2018ecoc} where a $96$-Gbaud signal is split into $S=7$
subbands. For a filter length of $13$, the MIMO filter in each SSFM
step can be represented as a $7 \times 7 \times 13$ tensor with $637$
real coefficients which is shown in Fig.~\ref{fig:subbands} (left).
The resulting complexity per step and subband would be almost $6$
times larger than that of the CD filters used in
\cite{Haeger2018ecoc}. The situation can be improved significantly by
decomposing each MIMO filter into a cascade of sparse filters as shown
in Fig.~\ref{fig:subbands} (right). For a cascade of $3$ filters, it
was shown that a simple $L_1$-norm regularization applied to the
filter coefficients during SGD leads to a sparsity level of round
$8$\%, i.e., $92$\% of the filter coefficients can be set to zero with
little performance penalty. Note that this filter decomposition
happens \emph{within} each SSFM step. In other words, complexity is
reduced by further increasing the depth of the multi-step DBP
computation graph. 

\subsection{Distributed PMD Compensation}
\label{sec:pmd}

\setcounter{figure}{2}
\begin{figure*}[t]
	\centering
	\includegraphics{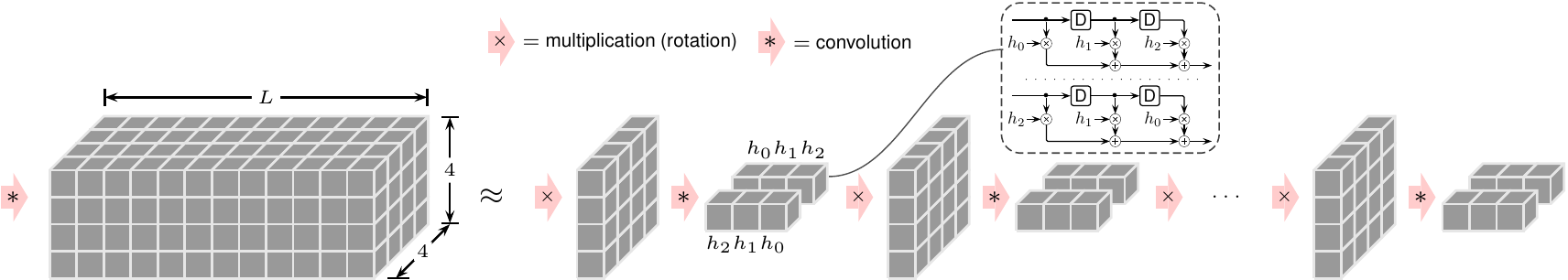}
%	\qquad
%	\subcaptionbox{}{\scalebox{0.85}{\includegraphics{../pstricks/fd_filter}}}
	\caption{Tensor representation of an $L$-tap $4 \times 4$ MIMO filter for
	PMD compensation (left); multi-step decomposition where
	$4$-D rotations are alternated with short fractional-delay (FD)
	filters accounting for DGD (right). Each FD filters is applied to
	both the real and imaginary part. }
	\label{fig:pmd_tensor}
\end{figure*}

Besides CD and fiber nonlinearity, PMD is another propagation
impairment that can be compensated using DSP. PMD is modeled by
dividing the fiber into $M$ sections, where the $i$-th section is
described by a frequency-dependent Jones matrix $\mat{R}^{(i)}
\mat{J}^{(i)}(\omega)$. Here, 
\begin{align}
	\label{eq:dgd}
	\mat{J}^{(i)}(\omega) = 
	\begin{pmatrix}
		e^{-\imag \omega \frac{\tau_i}{2}} & 0 \\
		0 & e^{\imag \omega \frac{\tau_i}{2}}  \\
	\end{pmatrix}
\end{align}
is a first-order PMD matrix with differential group delay (DGD)
$\tau_i$ and $\mat{R}^{(i)}$ is a complex unitary rotation matrix with
determinant one. In the linear regime, the effect of PMD for the
entire link is described by the matrix $\mat{J}(\omega) =
\prod_{i=1}^M \mat{R}^{(i)} \mat{J}^{(i)}(\omega)$ for sufficiently
large $M$. PMD compensation (and polarization demultiplexing) then
amounts to finding and applying the inverse $\mat{J}^{-1}(\omega)$ to
the received signal. 

Unlike CD, PMD is a time-varying effect and thus requires adaptive
compensation. In practical systems, adaptive PMD compensation is
typically performed after CD compensation, e.g., using an $L$-tap MIMO filter
that tries to approximate $\mat{J}^{-1}(\omega)$.
Fig.~\ref{fig:pmd_tensor} (left) shows the corresponding tensor
representation assuming a real-valued $4 \times 4$ filter that is
applied to the separated real and imaginary parts of both
polarizations \cite{Crivelli2014}. An efficient multi-step
decomposition of this filter is suggested in Fig.~\ref{fig:pmd_tensor}
(right), which essentially mimics the reverse propagation model by
alternating memoryless rotations and first-order PMD. Here, the
matrices \eqref{eq:dgd} are approximated with two real-valued
fractional-delay (FD) filters employing symmetrically flipped filter
coefficients for different polarizations. The FD filters can be very
short provided that the expected DGD per step is sufficiently small
(i.e., many steps are used). 

%(as opposed to a complex-valued $2 \times 2$ filter)

%(typically implemented via the constant modulus algorithms (CMA)
%\cite{Savory2008}) to an adaptive multi-stage PMD equalizer, similar
%to \cite{Liga2018}. 

Initial results suggest that the multi-step PMD architecture can be
effectively trained using SGD, similar to the conventional MIMO
equalizer. However, more research is needed to fully characterize the
training behavior, e.g., in terms of convergence speed for adaptive
compensation. The main advantage of the multi-step architecture is
that it can be easily integrated into DBP allowing for distributed PMD
compensation with low hardware complexity. Combining PMD compensation
with DBP has been previously studied for example in
\cite{Goroshko2016, Czegledi2017, Liga2018}. 

%Czegledi2016, 

\subsection{Coefficient Quantization and ASIC Implementation}

Fixed-point requirements and other DSP hardware implementation aspects
for DBP have been investigated in \cite{Fougstedt2017,
Fougstedt2018ecoc, Fougstedt2018ptl, Martins2018, Sherborne2018}. A
potential benefit of multi-step architectures is that they empirically
tend to have many ``good'' parameter configurations that lie
relatively close to each other. This implies that even if the
optimized parameters are slightly perturbed (e.g., by quantizing them)
there may exist a nearby parameter configuration that exhibits
similarly good performance to mitigate the resulting performance loss
due to the perturbation. 

Numerical evidence for this phenomenon can be obtained by considering
the joint optimization of CD filters in DBP including the effect of
filter coefficient quantization. This has been studied in
\cite{Fougstedt2018ecoc} and the approach relies on applying so-called
``fake'' quantizations to the filter coefficients, where the gradient
computations and parameter updates during SGD are still performed in
floating point. Compared to other quantization methods, this jointly
optimizes the responses of quantized filters and can lead to
significantly reduced fixed-point requirements. For the scenario in
\cite{Fougstedt2018ecoc}, it was shown for example that the bit
resolution can be reduced from $8$-$9$ coefficient bits to $5$--$6$
bits without adversely affecting performance. Furthermore, hardware
synthesis results in $28$-nm CMOS show that multi-step DBP based on TD
filtering with short FIR filters is well within the limits of current
ASIC technology in terms of chip area and power consumption
\cite{Fougstedt2018ecoc, Fougstedt2018ptl}.  

%\subsection{Mathematical equations}
%\begin{itemize}
%\item Equations should fit into a two-column print format and be single spaced.
%\item When writing mathematics, avoid confusion between characters that could be mistaken for one another, e.g. the letter `l' and the number one.
%\item Vectors and matrices should be in bold italic and variables in italic.
%\item If your paper contains superscripts or subscripts, take special care to ensure that the positioning of the characters is unambiguous.
%\item Exponential expressions should be written using superscript notation, i.e. $5\times 10^{3}$~not 5E03. 
%\item A multiplication symbol should be used, not a dot.
%\item Refer to equations using round brackets e.g. (1)
%\end{itemize}
%
%\subsection{Units}
%\begin{itemize}
%\item Use SI (MKS) units only and avoid spelling the unit in full instead of using the shortened notation e.g. use kJ not kilo Joules.
%\item If for any reason you must use mixed units, the units used for each quantity in an equation must be stated.
%\item Place a zero before decimal points: ``0.10'' do not put ``.10.'' 
%\end{itemize}

\section{Conclusion}

%Submissions should always include the following sections: an
%abstract; an introduction; a conclusion and a references section. If
%any of the above sections are not included the paper may be asked to
%add the relevant section or be rejected. 

We have illustrated through various examples how machine learning can
be used to achieve efficient fiber-nonlinearity compensation. Rather
than reducing the number of steps, it was highlighted that complexity
can also be reduced by carefully designing and optimizing multi-step
methods, or even by increasing the number of steps and decomposing
complex operations into simpler ones, without losing performance. We
also avoided the use of neural networks as universal (but sometimes
poorly understood) function approximators, and instead relied on
parameterizing existing model-based algorithms. 

\section{Acknowledgements}

%Acknowledgements should be placed after the conclusion and before the
%references section. Details of grants, financial aid and other special
%assistance should be noted.

\small This work is part of a project that has received funding from the
European Union's Horizon 2020 research and innovation programme under
the Marie Sk\l{}odowska-Curie grant agreement No.~749798. The work of
H.~D.~Pfister was supported in part by the National Science Foundation
(NSF) under Grant No.~1609327.  The work of A.~Alvarado and G.~Liga
has received funding from the European Research Council (ERC) under
the European Union's Horizon 2020 research and innovation programme
(grant agreement No.~757791).  Any opinions, findings,
recommendations, and conclusions expressed in this material are those
of the authors and do not necessarily reflect the views of these
sponsors.

%\section{References}

\normalsize

\clearpage 

\section{References}

\end{document}